\documentclass[1p]{elsarticle}

\usepackage{graphicx}

\newcommand{\be}{\begin{equation}}
\newcommand{\ee}{  \end{equation}}
\newcommand{\ba}{\begin{eqnarray}}
\newcommand{\ea}{  \end{eqnarray}}

\begin{document}

\begin{frontmatter}

\title{Doorway States in the Random-Phase Approximation}

\author[infn]{A. De Pace}
\author[dft,infn]{A. Molinari\fnref{*}}
\author[mpi]{H.A. Weidenm\"uller}

\address[infn]{Istituto Nazionale di Fisica Nucleare, Sezione di Torino, 
  via P.Giuria 1, I-10125 Torino, Italy}
\address[dft]{Dipartimento di Fisica Teorica dell'Universit\`a di Torino,
  via P.Giuria 1, I-10125 Torino, Italy}
\address[mpi]{Max-Planck-Institut f\"ur Kernphysik, D-69029 Heidelberg, Germany}
\fntext[*]{Deceased}

\begin{abstract}
  By coupling a doorway state to a sea of random background
  states, we develop the theory of doorway states in the framework of
  the random-phase approximation (RPA). Because of the symmetry of the
  RPA equations, that theory is radically different from the standard
  description of doorway states in the shell model. We derive the
  Pastur equation in the limit of large matrix dimension and show that
  the results agree with those of matrix diagonalization in large
  spaces. The complexity of the Pastur equation does not allow for an
  analytical approach that would approximately describe the doorway
  state. Our numerical results display unexpected features: The
  coupling of the doorway state with states of opposite energy leads
  to strong mutual attraction. 
\end{abstract}


\end{frontmatter}

\section{Introduction}
\label{int}

In the description of nuclear-structure phenomena, doorway states play
an important role. Standard examples are the giant-dipole
resonance~\cite{Har01} and, in medium-weight nuclei, low-lying
isobaric analogue states~\cite{Aue72}. A doorway state occurs when a
distinct mode of nuclear excitation of given spin and parity, coupled
strongly to the nuclear ground state or to some distinct scattering
channel, is mixed with a background of states with the same quantum
numbers. The strength of the mixing determines the spreading width of
the ensuing resonance. For the giant-dipole resonance in even-even
nuclei, the mode has spin/parity $1^-$ and is strongly coupled through
the dipole operator to the nuclear ground state. The background states
also have spin/parity $1^-$. The resonance shows up in the cross
section for photon absorption. The isobaric analogue state has isospin
$T_0 + 1$ and is strongly coupled to the channel for scattering of
protons on a target nucleus with one less proton. The background
states have isospin $T_0$. The isobaric analogue resonance shows up in
elastic proton scattering. Doorway states play an important role also
in other areas of physics.  By way of example we mention quantum
information theory, mesoscopic physics, quantum chaos, and molecular
physics. Without aiming at completeness, we refer to
Refs.~\cite{Nie00, Gor06, Abe08, Rid08, Jaq09, Shc09, Koh10a, Koh10b}
and references therein.

In the standard theoretical description (see Ref.~\cite{Wei09} and
references therein), the doorway mode has energy $E_0$ and is coupled
through real matrix elements $V_\mu$, $\mu = 1, \ldots, N$ to $N$
background states. These are governed by a real and symmetric
Hamiltonian matrix $h_{\mu \nu}$ with $\mu, \nu =1, \ldots, N$. In
matrix form the total Hamiltonian $H$ is given by
\be
H = \left( \matrix{ E_0 & V_\nu \cr
                   V_\mu & h_{\mu \nu} \cr} \right) \ .
\label{1}
\ee
(For isobaric analogue resonances, Eq.~(\ref{1}) must be generalized
to include the coupling of the analogue state with the background
states via the proton channel, see Ref.~\cite{Wei09}.) Eq.~(\ref{1})
is patterned after the nuclear shell model. There, the dipole mode
would be a linear superposition of one-particle one-hole states, the
background states would be two-particle two-hole states, and $E_0$,
the $V_\mu$ and the elements $h_{\mu \nu}$ would be determined by the
single-particle energies and the residual interaction. For $N \gg 1$ a
dynamical theory of the background states is not available in most
cases, however, and the Hamiltonian matrix $h$ is replaced by a matrix
drawn at random from the Gaussian Orthogonal Ensemble of real
symmetric matrices (the GOE). We have addressed the resulting problems
in the theory of doorway states in two recent papers. In
Ref.~\cite{Dep07}, we have worked out in a very general framework
properties of doorway states as averages over the GOE in the limit $N
\to \infty$. Properties of the spreading width that emerge beyond the
standard approximation were investigated in Ref.~\cite{Dep11}. An
essential and generic feature of the doorway state model is that the
value of the spreading width is adjustable. Moreover, this value is of
order $1 / N$ in relation to the overall width of the spectrum of the
background states. This last property guarantees that the doorway
state is a local spectral phenomenon.

In the present paper we extend the concept and the description of
doorway states to the random-phase approximation (RPA). Our extension
is motivated by the fact that in nuclear-structure theory, it is often
mandatory to replace the shell-model approach embodied in
Eq.~(\ref{1}) by the RPA~\cite{Rin80}. That is true especially for the
treatment of collective motion. The RPA is characterized by symmetries
that are radically different from those of the Hamiltonian approach in
Eq.~(\ref{1}). Our extension takes account of these symmetries.
Specifically, it involves four elements. (i) We need an RPA model for
the doorway state as a collective state. (ii) Similar to the
replacement of $h_{\mu \nu}$ by the GOE, our RPA model must involve a
random-matrix model with RPA symmetries for the background states.
(iii) The coupling of the doorway state to the background states (the
analogue of the matrix elements $V_\mu$) must also possess RPA
symmetries. (iv) The value of the spreading width due to that coupling
must be an adjustable parameter, and it must be of order $1 / N$ in
relation to the overall width of the spectrum of the background
states. The resulting theory of the doorway phenomenon in RPA turns
out to be radically different from the standard approach.

The random-matrix approach to RPA equations has been formulated and
investigated in some detail in Ref.~\cite{Bar09}. In that paper a
follow-up paper was announced that would combine the purely
statistical (or ``democratic'') description of the background states
in terms of a random-matrix model with the highly special dynamical
RPA description of a select state, the doorway state. Aside from being
an extension of our investigation of the doorway state phenomenon in
Refs.~\cite{Dep07, Dep11}, the present paper may also be viewed as
that follow-up paper. It might, therefore, also carry the title
``Random-Matrix Approach to RPA Equations II''. In the paper we are
mainly interested in the consequences the RPA symmetry has for the
doorway state picture. We do not discuss any applications.

\section{RPA}

For a set of $N$ degenerate shell-model states with energy $r$, the
RPA equations have the form~\cite{Rin80}
\be
{\cal H}^{(0)} X^\nu = (E_\nu - E_0) X^\nu
\label{2}
\ee
where
\be
{\cal H}^{(0)} = \left( \matrix{ r {\bf 1}_N + A & C \cr -
  C^* & - r {\bf 1}_N - A^* \cr} \right) \ .
\label{3}
\ee
In Eq.~(\ref{2}) $E_0$ ($E_\nu$) is the energy of the ground state
(one of the excited states, respectively). The eigenvectors $X^\nu$
have dimension $2 N$. In Eq.~(\ref{3}) ${\bf 1}_N$ is the unit matrix
in $N$ dimensions. The matrices $A$ and $C$ represent the residual
interaction and have dimension $N$ each. For time-reversal invariant
systems (orthogonal case), both $A$ and $C$ are real symmetric
matrices. If time reversal invariance is violated (unitary case), the
matrix $A$ is Hermitean and the matrix $C$ is complex symmetric. The
symmetries of the matrices $A$ and $C$ imply that if ${\cal E}_\nu =
E_\nu - E_0$ is an eigenvalue of the RPA equations~(\ref{2}), then
${\cal E}^*_\nu$, $- {\cal E}_\nu$, and $- {\cal E}^*_\nu$ are also
eigenvalues. These symmetries of ${\cal H}^{(0)}$ are caused by the
RPA approximation. As a consequence, the RPA matrix in Eq.~(\ref{3})
is not Hermitean and, therefore, does not correspond to any
Hamiltonian. That is why the random-matrix approach developed in
Ref.~\cite{Bar09} does not correspond to any of the generalized
random-matrix ensembles introduced by Altland and
Zirnbauer~\cite{Alt97}.

We now address point (i) of Section~\ref{int} and recall the
construction of a collective RPA state. That state will serve as the
doorway state. In the schematic model, such a collective state emerges
from the set of $N$ degenerate shell-model states if the interaction
is separable, i.e., if the RPA matrix in Eq.~(\ref{3}) is of the form
\be
\left( \matrix{ r \delta_{\mu \nu} + a_\mu a^*_\nu & a_\mu a_\rho \cr
  - a^*_\sigma a^*_\nu & - r \delta_{\sigma \rho} - a^*_\sigma a_\rho
  \cr} \right) \ .
\label{4}
\ee
The indices $\mu$ and $\nu$ ($\rho$ and $\sigma$) run from $1$ to $N$
(from $N + 1$ to $2 N$, respectively). The symmetry of the RPA
equations implies $a_\mu = a_{N + \mu}$ for all $\mu = 1, \ldots,
N$. The quantities $a_\mu$ with $\mu = 1, \ldots, N$ are complex in
the unitary case and real in the orthogonal case. In the schematic
model one often uses a common factor multiplying each of the four
separable matrices. The sign of that factor determines whether the
energy of the collective state is raised or lowered. We omit that
factor in order not to introduce too much complexity. The relative
strengths of the four separable matrices are fixed. This is neccessary
for the schematic model to yield a truly collective state. It is
straightforward to solve the RPA equations for the schematic
model~(\ref{4}). The collective state is located at energies $\pm [r^2
  + 2 r \sum_{\mu = 1}^N |a_\mu|^2]^{1/2}$ while the remaining $2(N -
1)$ states retain their unperturbed energies $\pm r$.

We turn to point (ii) of Section~\ref{int} and recall the
random-matrix approach to the RPA of Ref.~\cite{Bar09}. There it was
assumed that both $A$ and $C$ are random matrices. With the same
notation as used for the RPA matrix~(\ref{4}) we write the RPA
matrix~(\ref{3}) as
\be
\left( \matrix{ r \delta_{\mu \nu} + A_{\mu \nu} & C_{\mu \rho} \cr -
  C^*_{\sigma \nu} & - r \delta_{\sigma \rho} - A^*_{\sigma \rho} \cr} \right) \ .
\label{5}
\ee
The indices $\mu, \nu$ ($\rho, \sigma$) run from $1$ to $N$ (from $N +
1$ to $2 N$, respectively). Moreover we have $A_{(\mu + N) (\nu + N)}
= A_{\mu \nu}$ and $C_{\mu (\nu + N)} = C_{\mu \nu}$. The independent
elements of the matrices $A$ and $C$ are assumed to be uncorrelated
Gaussian--distributed random variables with zero mean values and
second moments given by
\ba
\langle A_{\mu \nu} A_{\mu' \nu'} \rangle &=& \frac{\lambda^2}{N}
\delta_{\mu \mu'} \delta_{\nu \nu'} \ {\rm (unitary \ case), \ or}
\nonumber \\
\langle A_{\mu \nu} A_{\mu' \nu'} \rangle &=& \frac{\lambda^2}{N}
(\delta_{\mu \mu'} \delta_{\nu \nu'} + \delta_{\mu \nu'} \delta_{\nu
\mu'}) \ {\rm (orthogonal \ case), \ and} \nonumber \\
\langle C_{\mu \nu} C^*_{\mu' \nu'} \rangle &=& \frac{\gamma^2}{N}
(\delta_{\mu \mu'} \delta_{\nu \nu'} + \delta_{\mu \nu'} \delta_{\mu'
\nu}) \ ; \ \langle C_{\mu \nu} C_{\mu' \nu'} \rangle = 0 \
{\rm (unitary \ case), \ or} \nonumber \\
\langle C_{\mu \nu} C_{\mu' \nu'} \rangle &=& \frac{\gamma^2}{N}
(\delta_{\mu \mu'} \delta_{\nu \nu'} + \delta_{\mu \nu'} \delta_{\nu
\mu'}) \ {\rm (orthogonal \ case)} \ .
\label{6}
\ea
In Eqs.~(\ref{6}) all indices run from $1$ to $N$. For $\gamma = 0$
and $r > 2 \lambda$ the average RPA spectrum consists of two
semicircles with equal radii $2 \lambda$ centered at $\pm r$.
Non-zero values of the parameter $\gamma$ cause attraction between
states with positive and negative energy and, with increasing
$\gamma$, eventually lead to coalescence of a first pair of levels
with opposite signs at energy $E = 0$. That is the point of
instability of the RPA equations. A further increase of $\gamma$ leads
to complex eigenvalues. In Ref.~\cite{Bar09} values of $\gamma$ at the
instability point are given as functions of the parameters $r$ and
$\lambda$.

Concerning point (iii) of Section~\ref{int} there are two alternative
ways of coupling the doorway state emerging from Eqs.~(\ref{2}),
(\ref{3}) and (\ref{4}) to the background states described by
Eqs.~(\ref{2}), (\ref{3}), (\ref{5}), and (\ref{6}).

\subsection{Strong-Coupling Model}
\label{str}

We combine the schematic model~(\ref{4}) for the doorway state with
the random RPA model of Eqs.~(\ref{5}) and (\ref{6}) and write the $2
N$--dimensional RPA matrix ${\cal H}$ as
\be
{\cal H} = \left( \matrix{ r \delta_{\mu
    \nu} + a_\mu a^*_\nu + A_{\mu \nu} & a_\mu a_\rho + C_{\mu \rho}
  \cr - a^*_\sigma a^*_\nu - C^*_{\sigma \nu} & - r \delta_{\rho
    \sigma} - a^*_\sigma a_\rho - A^*_{\sigma \rho} \cr} \right) \ .
\label{7}
\ee
The matrices $A$ and $C$ are taken to be members of the random-matrix
ensembles defined in and below Eqs.~(\ref{6}). Eq.~(\ref{7}) can be
interpreted by saying that the matrix elements of the two--body
interaction in the RPA approach partly factorize, giving rise to the
separable matrices $a_\mu a^*_\nu$ etc., while the remaining parts of
these matrix elements are replaced by random variables. That seems a
very natural choice of introducing both, a distinct doorway state and
a random-matrix description of the remaining background states.
Moreover, the strong-coupling model of Eq.~(\ref{7}) has the advantage
of carrying a small number of parameters. Unfortunately these
advantages are overcompensated by the fact that in the strong-coupling
model, the spreading width of the doorway state is fixed and given by
the radius $2 \lambda$ of the semicircle, i.e., of the average
spectrum of background states. We demonstrate that fact in the
appendix. Because of this shortcoming of the strong-coupling model we
devote the remainder of the paper entirely to the variable-coupling
model.

\subsection{Variable-Coupling Model}

In order to accommodate both, a set of $N$ random background states
and a collective state with variable coupling to these states, we
choose the RPA matrix ${\cal H}$ to have dimension $2N + 2$. We label
the first row and column with the letter $\alpha$, the following $N$
rows and columns with the indices $\mu, \nu = 1, \ldots, N$, the next
row and column with the letter $\beta$, and the remaining rows and
columns with the indices $\rho, \sigma = N + 1, \ldots, 2 N$.  We
write
\be
{\cal H} = \left( \matrix{ r + |a|^2 & v
  \delta_{\nu 1} & a^2 & w \delta_{\rho (N + 1)} \cr v^* \delta_{\mu
    1} & r \delta_{\mu \nu} + A_{\mu \nu} & w \delta_{\mu 1} & C_{\mu
    \rho} \cr - (a^*)^2 & - w^* \delta_{\nu 1} & - r - |a|^2 & - v^*
  \delta_{\rho (N + 1)} \cr - w^* \delta_{\sigma (N + 1)} & -
  C^*_{\sigma \nu} & - v \delta_{\sigma (N + 1)} & -r \delta_{\sigma
    \rho} - A^*_{\sigma \rho} \cr} \right) \ .
\label{10}
\ee
The matrices $A$ and $C$ are random, with the same properties as
described above. Comparison with expression~(\ref{4}) shows that the
collective state is already in (almost) diagonal form. In the RPA
matrix~(\ref{10}) the coupling to the background states is described
by the two parameters $v$ and $w$ that occur in the rows and columns
labelled $1$ and $N + 1$. All other coupling matrix elements vanish.
This simplified form of the RPA matrix~(\ref{10}) holds without loss
of generality. It follows from the generalized orthogonal (unitary)
invariance of the matrices $A$ and $C$. The parameters $v$ and $w$ are
real (complex) in the orthogonal (unitary) case. The form of the RPA
matrix~(\ref{10}) bears a direct analogy to the matrix on the
right-hand side of Eq.~(\ref{1}). There is some similarity between the
variable-coupling model of Eq.~(\ref{10}) and the strong-coupling
model of Eq.~(\ref{7}) except that in the latter the doorway state is
not afforded an extra dimension.

From the results of Ref.~\cite{Bar09} we expect that for $C = 0$,
coalescence of eigenvalues of the variable-coupling model occurs only
at $E = 0$ and only as a consequence of level repulsion caused by
$v$. To check this we put $C = 0$. Since $A$ causes level repulsion,
we also put $A = 0$. In the resulting eigenvalue equation, $N - 1$ of
the degenerate positive eigenvalues remain located at $r$ while one is
changed because of its interaction with the collective state. The same
is true of the negative eigenvalues. The relevant part of the secular
equation is
\be
\det \left( \matrix{ r + |a|^2 - E & v & a^2 & w \cr
  v^* & r - E & w & 0 \cr - (a^*)^2 & - w^* & - r - |a|^2 - E & - v^*
  \cr - w^* & 0 & - v & -r - E \cr} \right) = 0 \ .
\label{11}
\ee
The solution of this quadratic equation in $E^2$ is
\be
E^2 = r^2 + r |a|^2 + |v|^2 - |w|^2 \pm \sqrt{D} \ .
\label{12}
\ee
The discriminant $D$ has the value
\be
D = |a|^4 r^2 + 4 |v|^2 r^2 + 4 |v|^2 |a|^2 r - 2 r (a^2 v^* w^* +
{\rm c.c.}) \ .
\label{13}
\ee
Two eigenvalues coalesce when $D = 0$. For that to happen, the term
$(a^2 v^* w^* + {\rm c.c.})$ must be positive. With $\phi$ the phase
angle between $a^2 v^*$ and $w^*$ a zero of $D$ occurs if
\be
|w| \cos \phi = \frac{(1/4) r |a|^4 + |a|^2 |v|^2 + r |v|^2}{|a|^2 |v|}
\ .
\label{14}
\ee
Coalescence of two eigenvalues is physically relevant only if the
coalescing eigenvalues are real. That is the case if
\be
r^2 + r |a|^2 + |v|^2 - |w|^2 \geq 0 \ .
\label{15}
\ee
We use in Eq.~(\ref{15}) the equality sign, take $w, a, v$ to be
positive, and solve Eqs.~(\ref{14}) and (\ref{15}) for $v^2$. That
yields the two solutions $v^2_1 = (a^4/4) (r / (r + 2 a^2))$ and
$v^2_2 = a^4/4$. Insertion shows that for both $v^2_1$ and $v^2_2$,
the coinciding energy eigenvalues vanish. But for $v^2$ in the
interval $v^2_1 < v^2 < v^2_2$, the coinciding eigenvalues are real
and generically differ from zero. Choosing, for instance, $r = 2$ and
$a = 0.5$, one finds $v_1 = 0.111803$ and $v_2 = 0.125$ so that for
$v_1 < v < v_2$ the coinciding eigenvalues differ from zero. For
example, for $v = 0.12$ one gets $E = 0.113713$. The result is
somewhat surprising: In the variable-coupling model, coincidence of
two non-zero real eigenvalues is possible. This is in contrast to the
random RPA model in Eq.~(\ref{2}) where coincidence of two real
eigenvalues is possible only at $E = 0$. The appearance of coinciding
nonzero eigenvalues probably signals another breakdown point of the
RPA. Indeed, the values of $w$ corresponding to $v_1$ and $v_2$ are
$w_1 = 2.12426$ and $w_2 = 2.125$, respectively, much larger than
$v_1$ or $v_2$ and, thus, perhaps unphysical.

\section{Pastur Equation}
\label{pas}

We establish properties of the average spectrum of the
variable-coupling model with the help of the Pastur equation. We
follow Refs.~\cite{Dep07, Bar09, Dep11} and for simplicity consider
the unitary case only.

The Pastur equation is an equation for the average Green function, a
matrix of dimension $2 N + 2$ given by
\be
\langle G(E) \rangle =
\bigg\langle \bigg( E^+ {\bf 1}_{2 N + 2} - {\cal H} \bigg)^{-1}
\bigg\rangle \ ,
\label{16}
\ee
with ${\cal H}$ defined in Eq.~(\ref{10}). The energy $E$ carries a
positive imaginary increment. The angular brackets denote the ensemble
average. To obtain an equation for $\langle G(E) \rangle$, we define
the ``unperturbed'' Green function $G_0(E)$ by omitting from ${\cal H}$
the random matrices $A$ and $C$, and we expand $\langle G(E) \rangle$
around $G_0(E)$ in powers of $A$ and $C$. We take a term--by--term
ensemble average of the resulting series, omitting terms that are
small of order $1 / N$, and we resum the result.

The non-statistical part of ${\cal H}$ is given by
\be
{\cal H}_0 = \left( \matrix{
r + |a|^2 & v \delta_{\nu 1} & a^2 & w \delta_{\rho (N + 1)} \cr
v^* \delta_{\mu 1} & r \delta_{\mu \nu} & w \delta_{\mu 1} & 0 \cr
- (a^*)^2 & - w^* \delta_{\nu 1} & - r - |a|^2 & -
                                       v^* \delta_{\rho (N + 1)} \cr
- w^* \delta_{\sigma (N + 1)} & 0 & - v \delta_{\sigma (N + 1)} &
-r \delta_{\sigma \rho} \cr} \right) \ .  
\label{17}
\ee
The random part ${\cal H}_{\rm r}$ is correspondingly given by
\be
{\cal H}_{\rm r} = \left( \matrix{
0 & 0                  & 0 & 0                   \cr
0 & A_{\mu \nu}        & 0 & C_{\mu \rho}        \cr
0 & 0                  & 0 & 0                   \cr
0 & - C^*_{\sigma \nu} & 0 & - A^*_{\sigma \rho} \cr} \right) \ .
\label{18}
\ee
The unperturbed Green function is
\be
G_0(E) = (E^+ {\bf 1}_{2 N + 2} - {\cal H}_0 )^{-1} \ ,
\label{19}
\ee
and the Pastur equation reads
\be
\langle G(E) \rangle = G_0(E) + G_0(E) \langle {\cal H}_r \langle G(E)
\rangle {\cal H}_r \rangle \langle G(E) \rangle \ .
\label{20}
\ee
We define four projection operators, $Q_1$ and $Q_2$ as the projectors
onto the subspaces spanned by the states labelled $1,2,\ldots,N$ and
$N+1,N+2,\ldots,2N$, respectively, and $P_\alpha$ and $P_\beta$ as the
projectors onto the states $\alpha$ and $\beta$, respectively. For $i
= 1,2$ we define
\ba
\sigma_i(E) &=& \frac{\lambda}{N} {\rm Trace} \ Q_i \langle G(E) \rangle
Q_i \ , \nonumber \\ 
\delta &=& \frac{\gamma}{\lambda} \ , \nonumber \\
\Sigma_i(E) &=& \sigma_i - \delta^2 \sigma_{i + 1} \ , \nonumber \\
\sigma_3 &=& \sigma_1 \ .
\label{21}
\ea
Moreover we introduce
\ba
G^{(1)}_0(E) &=& Q_1 G_0(E)
+ Q_1 G_0(E) Q_2 \frac{\lambda \Sigma_2}{1 - \lambda \Sigma_2 Q_2
  G_0(E) Q_2} Q_2 G_0(E) \ , \nonumber \\ G^{(2)}_0(E) &=& Q_2 G_0(E)
+ Q_2 G_0(E) Q_1 \frac{\lambda \Sigma_1}{1 - \lambda \Sigma_1 Q_1
  G_0(E) Q_1} Q_1 G_0(E) \ .
\label{22}
\ea
Upon projection, the Pastur equation~(\ref{20}) yields the following
four equations.
\ba
P_\alpha \langle G(E) \rangle P_\alpha &=& P_\alpha G_0(E) P_\alpha
\nonumber \\
&& \qquad + \lambda \Sigma_1 P_\alpha G_0(E) Q_1 \bigg( 1 - \lambda
\Sigma_1 G^{(1)}_0(E) Q_1 \bigg)^{-1} G^{(1)}_0(E) P_\alpha \nonumber \\
&& \qquad + \lambda \Sigma_2 P_\alpha G_0(E) Q_2 \bigg( 1 - \lambda
\Sigma_2 G^{(2)}_0(E) Q_2 \bigg)^{-1} G^{(2)}_0(E) P_\alpha \ ,
\nonumber \\
P_\beta \langle G(E) \rangle P_\beta &=& P_\beta G_0(E) P_\beta \nonumber \\
&& \qquad + \lambda \Sigma_1 P_\beta G_0(E) Q_1 \bigg( 1 - \lambda
\Sigma_1 G^{(1)}_0(E) Q_1 \bigg)^{-1} G^{(1)}_0(E) P_\beta \nonumber \\
&& \qquad + \lambda \Sigma_2 P_\beta G_0(E) Q_2 \bigg( 1 - \lambda
\Sigma_2 G^{(2)}_0(E) Q_2 \bigg)^{-1} G^{(2)}_0(E) P_\beta \ ,
\nonumber \\
Q_1 \langle G(E) \rangle Q_1 &=& \bigg( 1 - \lambda \Sigma_1
G^{(1)}_0(E) Q_1 \bigg)^{-1} G^{(1)}_0(E) Q_1 \ , \nonumber \\
Q_2 \langle G(E) \rangle Q_2 &=& \bigg( 1 - \lambda \Sigma_2
G^{(2)}_0(E) Q_2 \bigg)^{-1} G^{(2)}_0(E) Q_2 \ .
\label{23}
\ea
Explicit calculation yields
\ba
(Q_1 G_0(E) Q_1)_{\mu \nu}     &=& \delta_{\mu \nu} (G_0(E))_{\mu \mu}
\ , \nonumber \\
(Q_2 G_0(E) Q_2)_{\rho \sigma} &=& \delta_{\rho \sigma} (G_0(E))_{\rho
\rho} \ , \nonumber \\
(Q_1 G_0(E) Q_2)_{\mu \rho}    &=& \delta_{\mu 1} \delta_{\rho (N+1)}
(G_0(E))_{1 (N+1)} \ , \nonumber \\
(Q_2 G_0(E) Q_1)_{\rho \mu}    &=& \delta_{\mu 1} \delta_{\rho (N+1)}
(G_0(E))_{(N+1) 1} \ , \nonumber \\
(P_\alpha G_0(E) Q_1)_{\alpha \mu} &=& \delta_{\mu 1} (G_0(E))_{\alpha 1} \ ,
\nonumber \\
(P_\alpha G_0(E) Q_2)_{\alpha \rho} &=& \delta_{\rho (N+1)} (G_0(E))_{\alpha
(N+1)} \ , \nonumber \\
(P_\beta G_0(E) Q_1)_{\beta \mu}   &=& \delta_{\mu 1} (G_0(E))_{\beta 1} \ ,
\nonumber \\
(P_\beta G_0(E) Q_2)_{\beta \rho}  &=& \delta_{\rho (N+1)} (G_0(E))_{\beta
(N+1)} \ , \nonumber \\
(Q_1 G_0(E) P_\alpha)_{\mu \alpha} &=& \delta_{\mu 1} (G_0(E))_{1 \alpha} \ ,
\nonumber \\
(Q_2 G_0(E) P_\alpha)_{\rho \alpha} &=& \delta_{\rho (N+1)} (G_0(E))_{(N+1)
\alpha} \ , \nonumber \\
(Q_1 G_0(E) P_\beta)_{\mu \beta}   &=& \delta_{\mu 1} (G_0(E))_{1 \beta} \ ,
\nonumber \\
(Q_2 G_0(E) P_\beta)_{\rho \beta}  &=& \delta_{\rho (N+1)} (G_0(E))_{(N+1)
\beta} \ .
\label{24}
\ea
We use Eqs.~(\ref{24}) in Eqs.~(\ref{22}) and (\ref{23}), define
\ba
(g_1)^{-1} &=& 1 - \lambda \Sigma_1 (G_0(E))_{1 1} \nonumber \\
&& - \lambda \Sigma_1 (G_0(E))_{1 (N+1)} \frac{\lambda \Sigma_2}{1 -
\lambda \Sigma_2 (G_0(E)_{(N+1) (N+1)}} (G_0(E))_{(N+1) 1} \ ,
\nonumber \\
(g_2)^{-1} &=& 1 - \lambda \Sigma_2 (G_0(E))_{(N+1) (N+1)}
\nonumber \\
&& - \lambda \Sigma_2 (G_0(E))_{(N+1) 1} \frac{\lambda \Sigma_1}{1 -
\lambda \Sigma_1 (G_0(E)_{1 1}} (G_0(E))_{1 (N+1)} \ ,
\label{25}
\ea
and obtain
\ba
&& (Q_1 \langle G(E) \rangle Q_1)_{\mu \nu} = \delta_{\mu \nu} (1 -
\delta_{\mu 1}) \frac{(G_0(E))_{\mu \mu}}{1 - \lambda \Sigma_1
(G_0(E))_{\mu \mu}} \nonumber \\
&& \qquad + \delta_{\mu \nu} \delta_{\mu
1} g_1 \bigg( (G_0(E))_{1 1} \nonumber \\
&& \qquad + (G_0(E))_{1 (N+1)} \frac{\lambda \Sigma_2}{1 - \lambda
\Sigma_2 (G_0(E))_{(N+1) (N+1)}}(G_0(E))_{(N+1) 1} \bigg) \ ,
\nonumber \\
&& (Q_2 \langle G(E) \rangle Q_2)_{\rho \sigma} = \delta_{\rho \sigma}
(1 - \delta_{\rho (N+1)}) \frac{(G_0(E))_{\rho \rho}} {1 - \lambda
\Sigma_2 (G_0(E))_{\rho \rho}} \nonumber \\
&& \qquad + \delta_{\rho \sigma} \delta_{\rho (N+1)} g_2 \bigg(
(G_0(E))_{(N+1) (N+1)} \nonumber \\
&& \qquad + (G_0(E))_{(N+1) 1} \frac{\lambda \Sigma_1}{1 - \lambda
\Sigma_1 (G_0(E))_{1 1}}(G_0(E))_{1 (N+1)} \bigg) \ , \nonumber \\ 
&& \langle G(E) \rangle_{\alpha \alpha} = (G_0(E))_{\alpha \alpha} +
\lambda \Sigma_1 (G_0(E))_{\alpha 1} g_1 \bigg( (G_0(E))_{1 \alpha}
\nonumber \\
&& \qquad + (G_0(E))_{1 (N+1)} \frac{\lambda \Sigma_2}{1 - \lambda
\Sigma_2 (G_0(E))_{(N+1) (N+1)}} (G_0(E))_{(N+1) \alpha} \bigg)
\nonumber \\
&& \qquad + \lambda \Sigma_2 (G_0(E))_{\alpha (N+1)} g_2 \bigg(
(G_0(E))_{(N+1) \alpha} \nonumber \\
&& \qquad + (G_0(E))_{(N+1) 1} \frac{\lambda \Sigma_1}{1 - \lambda
\Sigma_1 (G_0(E))_{1 1}} (G_0(E))_{1 \alpha} \bigg) \ , 
\nonumber \\
&& \langle G(E) \rangle_{\beta \beta} = (G_0(E))_{\beta \beta} +
\lambda \Sigma_1 (G_0(E))_{\beta 1} g_1 \bigg( (G_0(E))_{1 \beta}
\nonumber \\
&& \qquad + (G_0(E))_{1 (N+1)} \frac{\lambda \Sigma_2}{1 - \lambda
\Sigma_2 (G_0(E))_{(N+1) (N+1)}} (G_0(E))_{(N+1) \beta} \bigg)
\nonumber \\
&& \qquad + \lambda \Sigma_2 (G_0(E))_{\beta (N+1)} g_2 \bigg(
(G_0(E))_{(N+1) \beta} \nonumber \\
&& \qquad + (G_0(E))_{(N+1) 1} \frac{\lambda \Sigma_1}{1 - \lambda
\Sigma_1 (G_0(E))_{1 1}} (G_0(E))_{1 \beta} \bigg) \ . 
\label{26}
\ea
We take the trace of the first and the second of these equations and
use the definitions~(\ref{21}). That yields
\ba
\sigma_1(E) &=& \frac{N-1}{N} \frac{\lambda}{E - r - \lambda
\Sigma_1} + \frac{1}{N} \lambda g_1 \bigg( (G_0(E))_{1 1}
\nonumber \\
&& \qquad + (G_0(E))_{1 (N+1)} \frac{\lambda \Sigma_2}{1 - \lambda
\Sigma_2 (G_0(E))_{(N+1) (N+1)}} (G_0(E))_{(N+1) 1} \bigg) \ , 
\nonumber \\
\sigma_2(E) &=& \frac{N-1}{N} \frac{\lambda}{E + r - \lambda
\Sigma_2} + \frac{1}{N} \lambda g_2 \bigg( (G_0(E))_{(N+1) (N+1)}
\nonumber \\
&& \qquad + (G_0(E))_{(N+1) 1} \frac{\lambda \Sigma_1}{1 - \lambda
\Sigma_1 (G_0(E))_{1 1}} (G_0(E))_{1 (N+1)} \bigg) \ .
\label{27}
\ea
Given the parameters $r, a, v, w$ we can calculate the matrix elements
$(G_0(E))_{i j}$ with $i, j = \alpha, 1, \beta, (N+1)$. Then
Eqs.~(\ref{27}) together with the defining Eqs.~(\ref{25}) constitute
a pair of non-linear coupled equations for the unknown functions
$\sigma_1(E)$ and $\sigma_2(E)$. Once these are known, $\langle G(E)
\rangle_{\alpha \alpha}$ and $\langle G(E) \rangle_{\beta \beta}$ are
given by the last two Eqs.~(\ref{26}). The symmetry properties of the
solutions of Eqs.~(\ref{27}) are the same as discussed in Section~6 of
Ref.~\cite{Bar09}.

\begin{figure}[t]
\begin{center}
\includegraphics[clip,width=0.9\textwidth]{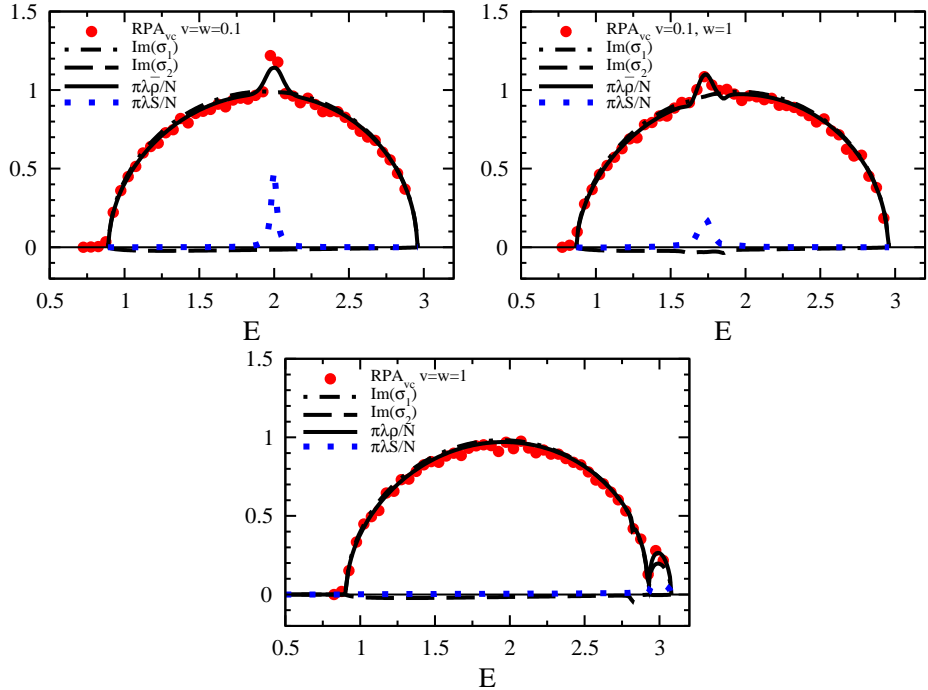}
\caption{The dimensionless total level density as obtained from the
  Pastur equation (black line) and from matrix diagonalization (red
  dots) plus various contributions to the density (see text) for three
  values of $(v, w)$ as indicated in the figure.}
\label{fig1}
\end{center}
\end{figure}

To determine the average level density $\rho(E) = \rho(-E)$, we focus
attention on the positive part of the spectrum. We expect that part to
consist of up to three pieces, one due to the random part of ${\cal
  H}$, one due to the level with index $\alpha$, and one due to the level
with index $1$. (That last level interacts strongly with level
$\alpha$ and may be pushed outside the random part of the spectrum).
The average Green function has up to three branch cuts along the
positive real $E$ axis, each corresponding to one of these parts. We
look for the solutions with negative imaginary parts on each of the
cuts. For the solutions with $E < 0$ we refer to the discussion in
Section~6 of Ref.~\cite{Bar09}. For $E > 0$ the level density is
given by
\be
\rho(E) = - \frac{N}{\pi \lambda} \Im \big( \sigma_1(E) +
\sigma_2(E) \big) + S(E) \ .
\label{28}
\ee
Here we allow for the possibility that the imaginary parts of
$\sigma_2$ and of $\langle G(E) \rangle_{\beta \beta}$ do not vanish
for $E > 0$. With $\langle G_{\alpha \alpha}(E) \rangle$ and $\langle
G_{\beta \beta}(E) \rangle$ given by the last two Eqs.~(\ref{26}), the
strength function is
\be
S(E) = - \frac{1}{\pi} \Im [ \langle G_{\alpha \alpha}(E) \rangle
+ \langle G_{\beta \beta}(E) \rangle ] \ .
\label{29}
\ee
The doorway state is expected to lead to a local enhancement of
$\rho(E)$ caused by the strength function $S(E)$.

\section{Numerical Results}

\begin{figure}[t]
\begin{center}
\includegraphics[clip,width=0.9\textwidth]{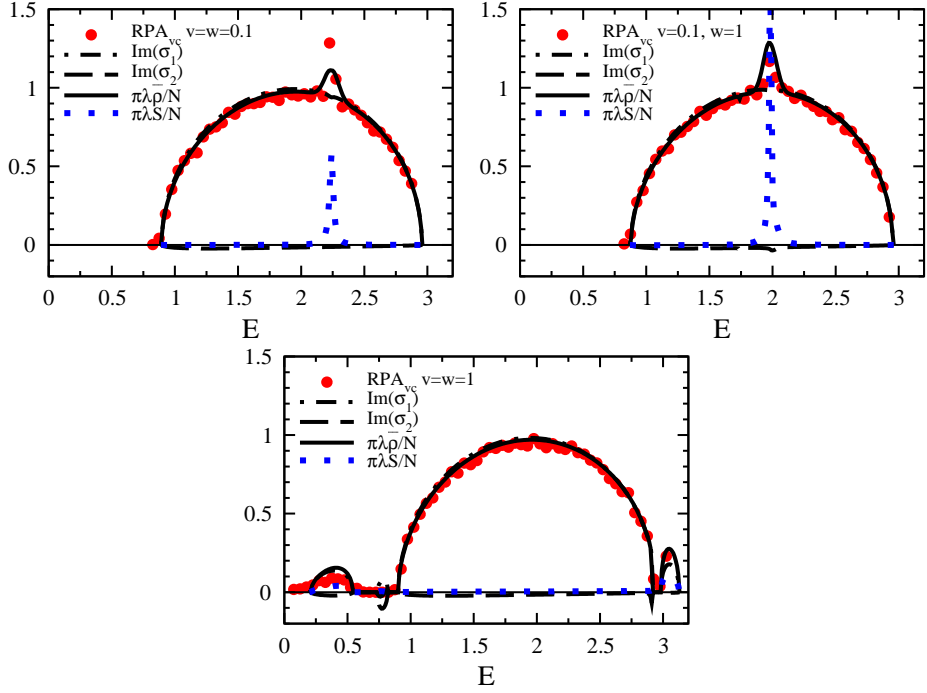}
\caption{Same as Figure~\ref{fig1} but for $a = 0.5$.}
\label{fig2}
\end{center}
\end{figure}

In Refs.~\cite{Dep07, Dep11} we have given explicit analytical
expressions for the location and the spreading width of the doorway
state as functions of the parameters of the underlying dynamical
model. In spite of determined efforts we have not been able to derive
similarly useful expressions for the RPA approach. The reason is that
in comparison to the shell-model case of Eq.~(\ref{1}), the matrix
dimension of the RPA approach of Eq.~(\ref{10}) is doubled. In the
shell-model case of Eq.~(\ref{1}), the Pastur equation is of second
order and lends itself to a straightforward analytical treatment. In
contradistinction, the Pastur equations~(\ref{27}) involve two
functions $\sigma_1(E)$ and $\sigma_2(E)$ and are effectively of
fourth order. Even within a perturbative treatment of the doorway
state, the complexity of the resulting expressions for $\rho(E)$ and
$S(E)$ is such that they do not elucidate the physical properties of
the doorway state. For these reasons we confine ourselves to a
presentation and discussion of the numerical results.

\begin{figure}[t]
\begin{center}
\includegraphics[clip,width=0.9\textwidth]{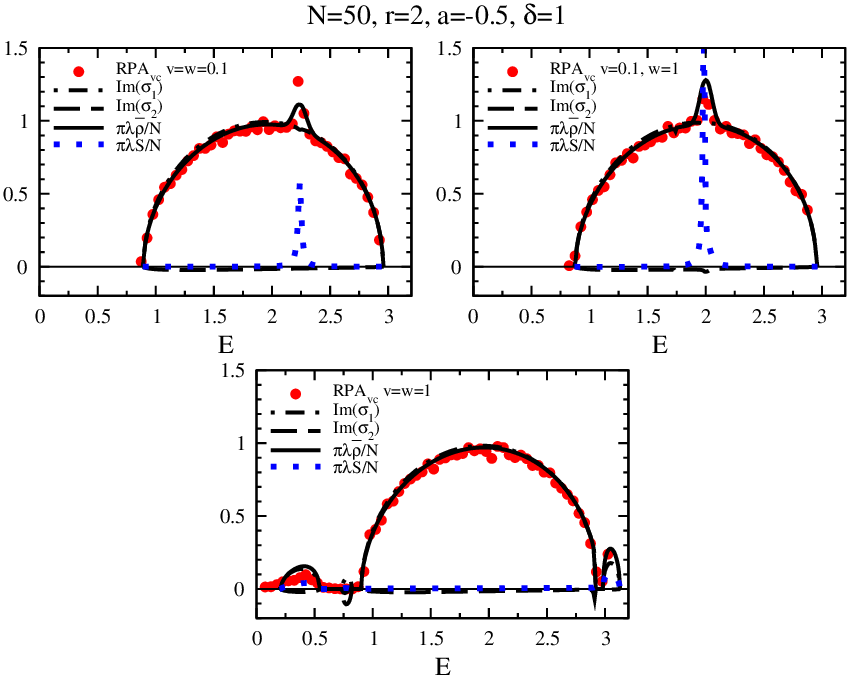}
\caption{Same as Figure~\ref{fig1} but for $a = - 0.5$.}
\label{fig3}
\end{center}
\end{figure}

The parameters of the variable-coupling model were chosen as follows.
The location of the unperturbed background states was taken at $r =
2$, see Eq.~(\ref{4}). The dimension of the random matrices $A$ and
$C$ in Eq.~(\ref{5}) was $N = 50$, the half width of the spectrum of
the random matrix $A$ was chosen as $\lambda = 1/2$, see
Eqs.~(\ref{6}). These parameters were held fixed for all cases
calculated. Without coupling to the doorway state and for $C = 0$, the
spectrum of $A$ extends from $E = 1$ to $E = 3$ and from $E = - 3$ to
$E = - 1$. We have done calculations for several values of the
parameter $a$ (the unperturbed location of the doorway state in
Eq.~(\ref{10})), $v$ and $w$ (strength of the coupling between the
doorway state and the background states in Eq.~(\ref{10})), and
$\delta$ (relative strength of the random matrices $C$ and $A$ as
defined in Eqs.~(\ref{21})).  In the figures shown we confine
ourselves to values of $a$ such that the unperturbed position of the
doorway state is within the spectrum of background states generated by
$A$.  Only the positive part of the energy spectrum is shown in all
figures. 

In the case of the numerical diagonalization of the RPA equations, the level
density is calculated as number of states per bin. In order to limit
statistical fluctuations, the bin width cannot be taken arbitrarily small and
it turns out to be typically larger than the doorway state width. Hence, to
make a meaningful comparison between the numerical diagonalization and the
Pastur results, we have introduced a smoothed strength function $\bar{S}(E)$,
obtained by convoluting the strength function (\ref{29}) with a Gaussian
distribution of variance equal to the bin width. For the purpose of comparison, 
in the figures we display the smoothed dimensionless total level density 
$\pi\lambda\bar{\rho}(E)/N$, with $\bar{\rho}$ defined as 
\be
\bar{\rho}(E) = - \frac{N}{\pi \lambda} \Im \big( \sigma_1(E) +
\sigma_2(E) \big) + \bar{S}(E) \ ,
\label{30}
\ee
instead of $\rho$ as given in Eq.~(\ref{28}).

In the figures we also display the contributions due to $- \Im \sigma_1$ and 
$- \Im \sigma_2$ as defined in Eqs.~(\ref{27}) and to $\pi \lambda S(E) / N$ 
with $S(E)$ defined in Eq.~(\ref{29}) (i.~e., the strength function as obtained
from the Pastur equations). The result of the numerical diagonalization
of the RPA equations is shown as red dots.

\begin{figure}[t]
\begin{center}
\includegraphics[clip,width=0.9\textwidth]{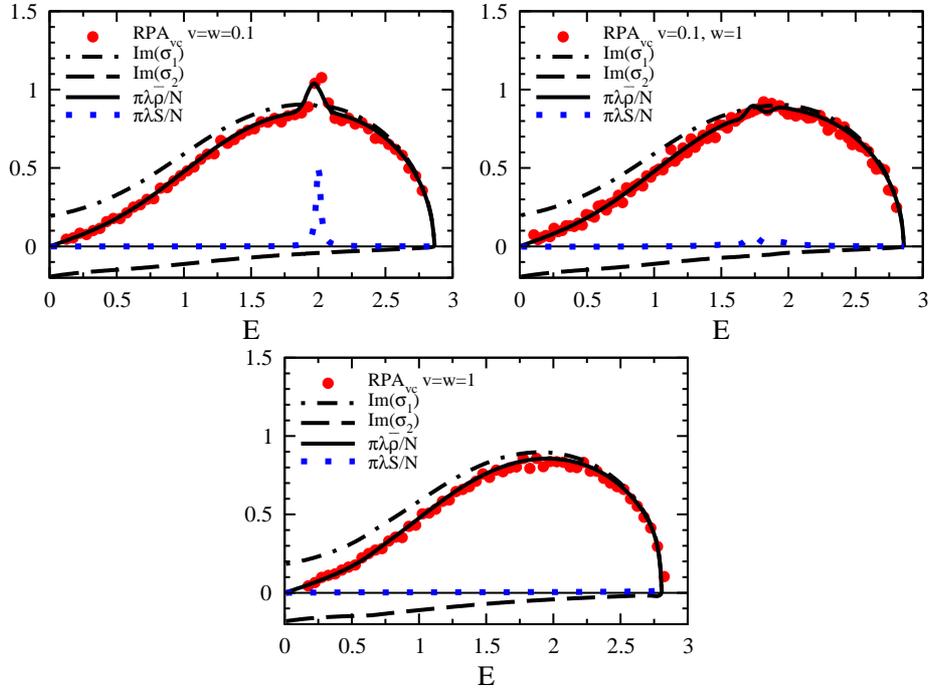}
\caption{Same as Figure~\ref{fig1} but for changed values of the
  parameters as indicated.}
\label{fig4}
\end{center}
\end{figure}

The top left panel of Fig.~\ref{fig1}, obtained for weak coupling, shows
the expected pattern of a doorway state in the center of and
superposed over the semicircular spectrum of background states. Due to
the coupling matrix $C$ that spectrum is displaced toward the left
from its original position for $C = 0$. As the strength parameter $w$
is increased (second panel), the doorway state is shifted toward
smaller energies. This reflects the increased attraction between
states of positive and negative energy already observed in
Ref.~\cite{Bar09}. Such attraction is due to terms like $w$ or $C$
that connect the positive- and the negative-energy parts of the RPA
matrix. Keeping the parameter $w$ fixed and increasing the parameter
$v$ (third panel) causes level repulsion and shifts the doorway state
outside the spectrum of background states. The same pattern is
discernible in Figs.~\ref{fig2} and \ref{fig3} where the original
location of the doorway state is defined by $a = 0.5$ and by $a = -
0.5$, respectively. In all three figures there appear bumps at small
energies for $v=w = 1$. We have not investigated these more closely.

\begin{figure}[t]
\begin{center}
\includegraphics[clip,width=0.9\textwidth]{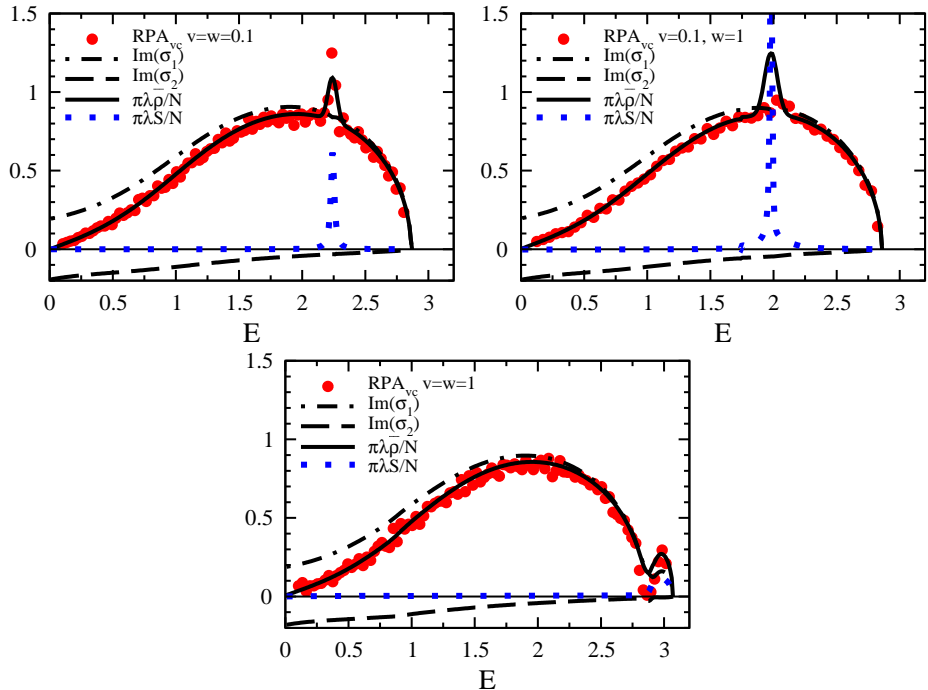}
\caption{Same as Figure~\ref{fig1} but for changed values of the
  parameters as indicated.}
\label{fig5}
\end{center}
\end{figure}

Figures~\ref{fig4} and \ref{fig5} show cases where the coupling
$\delta$ between positive- and negative-energy states is increased to
$\delta = 2$. In that case, attraction between the two parts of the
spectrum significantly shifts and distorts the semicircular spectrum
of the background states, see Ref.~\cite{Bar09}. For $a = 0$
(Fig.~\ref{fig4}) and weak coupling (top left panel) the doorway state is
unaffected by this change. However, it completely loses its identity
and disappears among the background states when either $w$ or both $v$
and $w$ are significantly increased. That is not the case for $a =
0.5$ where the doorway state keeps its identity but moves in a manner
similar to that of Figures~\ref{fig1} to \ref{fig3}. The difference is
due to the contribution of the doorway state shown in
Figures~\ref{fig4} and \ref{fig5}.

\begin{figure}[t]
\begin{center}
\includegraphics[clip,width=0.9\textwidth]{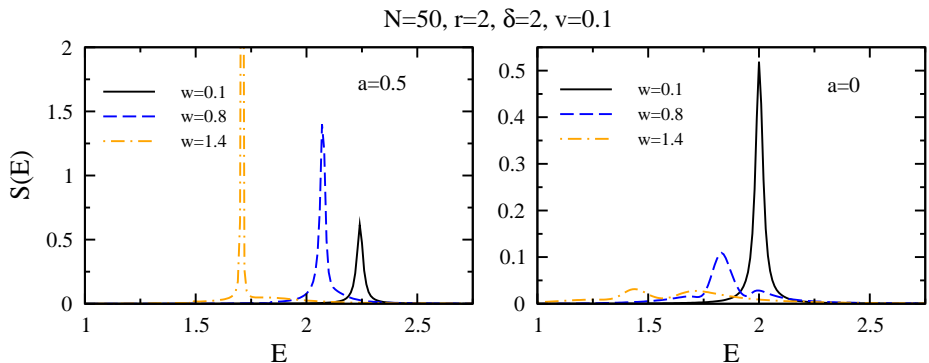}
\caption{The evolution of the doorway strength function for two different state
position varying the strength of the coupling.}
\label{fig6}
\end{center}
\end{figure}

In Fig.~\ref{fig6} we display the strikingly different behavior of the
doorway state strength function when the position $a$ of the collective state
moves out of the background center. For $a=0$ the strength decreases when the
coupling to the negative energy states is increased, whereas the opposite
happens when the collective state is not centered. This behavior is the reason 
for the differences between the top right panels of Figures~\ref{fig4} and
\ref{fig5}. 

Finally, note that the doorway strength is of order $1/N$ with respect to the
background (see Eq.~(\ref{28})), so that for values of $N$ larger than the one
employed in the figures ($N=50$) it would become progressively less visible,
decreasing its heigth as $1/N$.

\section{Conclusions}

In the framework of the RPA approximation we have investigated a
doorway state coupled to a sea of $N$ background states. The latter
are described in terms of the random-matrix model for the RPA
equations developed and investigated in Ref.~\cite{Bar09}. The
symmetry of the RPA equations allows for two possibilities, strong
coupling or variable coupling of the doorway state to the background
states. We have shown that the first alternative is physically not
interesting, and we have not considered it in the paper. For the
second alternative we have in the limit $N \to \infty$ derived the
Pastur equation for the average level density. That equation has twice
the dimension of the standard case. In contrast to the standard
situation, that makes it virtually impossible to obtain even
approximate expressions for the location and width of the doorway
state. Therefore, we have confined ourselves to a numerical approach.
We have shown that the solution of the Pastur equation agrees well
with the results of matrix diagonalization for $N = 50$.

We have shown that the theoretical description of doorway states
within the RPA framework is distinctly different from the standard
treatment. The two interaction matrix elements $v$ and $w$ that
characterize the interaction of the doorway state with the background
states play very different roles. While $v$ couples the doorway state
to states of the same (positive or negative) energy, $w$ couples the
doorway state at positive energy to background states at negative
energy and vice versa. The matrix element $v$ causes a spreading of
the doorway state and the associated level repulsion, similar to the
standard situation. The matrix element $w$ causes level attraction and
moves the position of the doorway state rather strongly without
discernibly affecting its width.

{\it Acknowledgements.} One of us (HAW) is grateful to the late
O. Bohigas for discussions that sparked the present investigation.

\section*{Appendix: Failure of the Strong-Coupling Model}

We show that in the strong-coupling model, the spreading width is of
the order of the total width of the spectrum. To that end we display
the doorway state in the matrix ${\cal H}$ explicitly. The separable
matrix with elements $a_\mu a^*_\nu$ in Eq.~(\ref{7}) has a single
non-vanishing eigenvalue $|a|^2$ and is diagonalized by a unitary
matrix $U$ that obeys $\sum_\nu U_{\mu \nu} a_\nu = \delta_{\mu 1}
a$. We use the transformation
\ba
{\cal H} &\to& \left(
\matrix{ U & 0 \cr 0 & U^* \cr} \right) {\cal H} \left( \matrix{
  U^\dag & 0 \cr 0 & U^T \cr} \right) \nonumber \\ &=& \left( \matrix{
  r \delta_{\mu \nu} & 0 \cr 0 & -r \delta_{\mu \nu} \cr} \right)
\nonumber \\ &+& \left( \matrix{ |a|^2 \delta_{\mu 1} \delta_{\nu 1} &
  a^2 \delta_{\mu 1} \delta_{\rho (N + 1)} \cr - (a^*)^2
  \delta_{\sigma (N + 1)} \delta_{\nu 1} & - |a|^2 \delta_{\sigma (N +
    1)} \delta_{\rho (N + 1)} \cr} \right) \nonumber \\ &+& \left(
\matrix{ (U A U^\dag)_{\mu \nu} & (U C U^T)_{\mu \rho} \cr - (U^* C^*
  U^\dag)_{\sigma \nu} & - (U^* A^* U^T)_{\sigma \rho} \cr} \right)
\ .
\label{8}
\ea
Because of generalized unitary invariance, the transformed matrices $U
A U^\dag$ and $U C U^T$ belong to the same ensemble as the matrices
$A$ and $C$, respectively, and we replace them by the latter without
loss of generality. Then the sum of the first and the third terms on
the right-hand side of Eq.~(\ref{8}) yields ${\cal H}^{(0)}$ in
Eq.~(\ref{3}), and we have
\be
{\cal H} = {\cal H}^{(0)} +
\left( \matrix{ |a|^2 \delta_{\mu 1} \delta_{\nu 1} & a^2 \delta_{\mu
    1} \delta_{\rho (N + 1)} \cr - (a^*)^2 \delta_{\sigma (N + 1)}
  \delta_{\nu 1} & - |a|^2 \delta_{\sigma (N + 1)} \delta_{\rho (N +
    1)} \cr} \right) \ .
\label{9}
\ee
The strong-coupling model has four parameters, the three parameters
$r, \lambda, \gamma$ of the stochastic model and the parameter $a$
that determines the position of the collective state. There is no
independent parameter that would determine the strength of the
coupling between the collective state and the background states.
According to Eq.~(\ref{9}) that coupling is mediated by those matrix
elements of $A$ and of $C$ that appear in the first and the $N+1$st
rows and columns of ${\cal H}^{(0)}$. For a semiquantitative estimate
we apply the standard expression $\Gamma^\downarrow = 2 \pi \langle
|v|^2 \rangle \rho(E)$ for the spreading width to our case. Here
$\langle |v|^2 \rangle$ is the mean square matrix element coupling the
doorway state and the background states, and $\rho(E)$ is the average
level density of the latter. In the present case we have $\langle
|v|^2 \rangle = \langle |A|^2_{1 1} \rangle = \lambda^2 / N$ and, in
the center of the GOE spectrum, $\rho(E) = N / (\pi \lambda)$. Thus,
$\Gamma^\downarrow = 2 \lambda$ equals the width of the spectrum of
the background states. That feature violates postulate (iv) of
Section~\ref{int} and is totally unphysical. It implies that the
resonance due to the doorway state cannot be distinguished from the
spectrum of the background states: There simply is no identifiable
doorway state. A more detailed investigation that takes account of the
matrices $C$ via the Pastur equation does not modify that conclusion
in any essential way since there is no mechanism to reduce the value
of the spreading width by a factor $1 / N$.

\end{document}